\documentclass[conference]{IEEEtran}
\makeatletter\if@twocolumn\PassOptionsToPackage{switch}{lineno}\else\fi\makeatother

\usepackage{graphicx}
\usepackage[T1]{fontenc}

\usepackage{booktabs}
\usepackage{tabu,multirow}
\usepackage[table,xcdraw]{xcolor}
\usepackage{amsmath}
\usepackage{subfigure} 

\ifCLASSINFOpdf
\else
\fi
\usepackage{amsmath}
\usepackage[cmintegrals]{newtxmath}
\usepackage{url,multirow,morefloats,floatflt,cancel,tfrupee}
\makeatletter

\AtBeginDocument{\@ifpackageloaded{textcomp}{}{\usepackage{textcomp}}}
\makeatother
\usepackage{colortbl}
\usepackage{xcolor}
\usepackage{pifont}
\usepackage[nointegrals]{wasysym}
\makeatletter

\def\mcWidth#1{\csname TY@F#1\endcsname+\tabcolsep}

\def\cAlignHack{\rightskip\@flushglue\leftskip\@flushglue\parindent\z@\parfillskip\z@skip}
\def\rAlignHack{\rightskip\z@skip\leftskip\@flushglue \parindent\z@\parfillskip\z@skip}

\@ifundefined{etal}{}{}

\usepackage{ifxetex}
\ifxetex\else\if@twocolumn\@ifpackageloaded{stfloats}{}{\usepackage{dblfloatfix}}\fi\fi

\AtBeginDocument{
\expandafter\ifx\csname eqalign\endcsname\relax
\def\eqalign#1{\null\vcenter{\def\\{\cr}\openup\jot\m@th
  \ialign{\strut$\displaystyle{##}$\hfil&$\displaystyle{{}##}$\hfil
      \crcr#1\crcr}}\,}
\fi
}

\AtBeginDocument{%
  \@ifpackageloaded{endfloat}%
   {\renewcommand\efloat@iwrite[1]{\immediate\expandafter\protected@write\csname efloat@post#1\endcsname{}}}{\newif\ifefloat@tables}%
}%

\def\BreakURLText#1{\@tfor\brk@tempa:=#1\do{\brk@tempa\hskip0pt}}
\let\lt=<
\let\gt=>
\def\processVert{\ifmmode|\else\textbar\fi}

\@ifundefined{subparagraph}{
\def\subparagraph{\@startsection{paragraph}{5}{2\parindent}{0ex plus 0.1ex minus 0.1ex}%
{0ex}{\normalfont\small\itshape}}%
}{}

\newcommand\role[1]{\unskip}
\newcommand\aucollab[1]{\unskip}
  
\@ifundefined{tsGraphicsScaleX}{\gdef\tsGraphicsScaleX{1}}{}
\@ifundefined{tsGraphicsScaleY}{\gdef\tsGraphicsScaleY{.9}}{}
\def\checkGraphicsWidth{\ifdim\Gin@nat@width>\linewidth
	\tsGraphicsScaleX\linewidth\else\Gin@nat@width\fi}

\def\checkGraphicsHeight{\ifdim\Gin@nat@height>.9\textheight
	\tsGraphicsScaleY\textheight\else\Gin@nat@height\fi}

\def\fixFloatSize#1{}
\let\ts@includegraphics\includegraphics

\def\inlinegraphic[#1]#2{{\edef\@tempa{#1}\edef\baseline@shift{\ifx\@tempa\@empty0\else#1\fi}\edef\tempZ{\the\numexpr(\numexpr(\baseline@shift*\f@size/100))}\protect\raisebox{\tempZ pt}{\ts@includegraphics{#2}}}}

\AtBeginDocument{\def\includegraphics{\@ifnextchar[{\ts@includegraphics}{\ts@includegraphics[width=\checkGraphicsWidth,height=\checkGraphicsHeight,keepaspectratio]}}}

\DeclareMathAlphabet{\mathpzc}{OT1}{pzc}{m}{it}

\def\URL#1#2{\@ifundefined{href}{#2}{\href{#1}{#2}}}

\def\UrlOrds{\do\*\do\-\do\~\do\'\do\"\do\-}%
\g@addto@macro{\UrlBreaks}{\UrlOrds}

\edef\fntEncoding{\f@encoding}

\makeatother

\newif\ifmultipleabstract\multipleabstractfalse%
\makeatletter
\AtBeginDocument{\@ifpackageloaded{longtable}{%
\def\LT@makecaption#1#2#3{%
  \LT@mcol\LT@cols c{\hbox to\z@{\hss\parbox[t]\LTcapwidth{%
    \sbox\@tempboxa{#1{#2: } #3}%
    \ifdim\wd\@tempboxa>\hsize
      #1{#2: }\textsc{#3}%
    \else
      \hbox to\hsize{\hfil\box\@tempboxa\hfil}%
    \fi
    \endgraf\vskip\baselineskip}%
  \hss}}}
}{}}
\makeatother

\begin{document}

%


        \title{Attention-Guided Version of 2D UNet for Automatic Brain Tumor Segmentation}
           \author{
		\IEEEauthorblockN{Mehrdad~Noori, Ali~Bahri, Karim~Mohammadi}

    \IEEEauthorblockA{School of Electrical Engineering, \\Iran University of Science and Technology,\\ 
        Tehran, 
        Iran}\\[-12pt]me.noori.1994@gmail.com, alibahri.72.dl.k@gmail.com, mohammadi@iust.ac.ir }



\maketitle 

\IEEEpubidadjcol


\begin{abstract}
Gliomas are the most common and aggressive among brain tumors, which cause a short life expectancy in their highest grade. Therefore, treatment assessment is a key stage to enhance the quality of the patients' lives. Recently, deep convolutional neural networks (DCNNs) have achieved a remarkable performance in brain tumor segmentation, but this task is still difficult owing to high varying intensity and appearance of gliomas. Most of the existing methods, especially UNet-based networks, integrate low-level and high-level features in a naive way, which may result in confusion for the model. Moreover, most approaches employ 3D architectures to benefit from 3D contextual information of input images. These architectures contain more parameters and computational complexity than 2D architectures. On the other hand, using 2D models causes not to benefit from 3D contextual information of input images. In order to address the mentioned issues, we design a low-parameter network based on 2D UNet in which we employ two techniques. The first technique is an attention mechanism, which is adopted after concatenation of low-level and high-level features. This technique prevents confusion for the model by weighting each of the channels adaptively. The second technique is the Multi-View Fusion. By adopting this technique, we can benefit from 3D contextual information of input images despite using a 2D model. Experimental results demonstrate that our method performs favorably against 2017 and 2018 state-of-the-art methods. \end{abstract}



%
\IEEEpeerreviewmaketitle

\section{Introduction}
Gliomas are the most prevalent brain tumors which occur more frequently in adults and may be initiated by glial cells~\cite{mamelak2007targeted}. They account for nearly eighty percent of all malignant brain tumors diagnosed in the United States~\cite{gholami2016inverse}. There are two types of Glioma including High-grade Glioma (HGG) and Low-grade Glioma (LGG). HGG tumors are malignant and grow quickly that usually require a surgery where the average survival period for affected patients has been reported two years or less. LGG tumors are followed by a few years of life expectancy along with the aggressive treatment being sometimes delayed as much as possible~\cite{saut2014multilayer}. 

To analyze and monitor brain tumor images, there are some major tools such as Magnetic Resonance Imaging (MRI). The MRI provides detailed brain images and is a common imaging method which is utilized to visualize the extent of tumor regions. However, the manual segmentation of 3D MRI scans requires significant amount of time and is susceptible to inaccuracies and variability due to the highly complex nature of tumor appearance. Accordingly, automatic brain tumor segmentation of MRI images can dramatically affect the improvement of diagnosis, prediction of growth rates, and treatment plans, particularly where access to an expert radiologist is restricted.

A brain is visualized by employing different MRI modalities. The most commonly used MRI modalities for brain tumor segmentation are as follows: T1-weighted, post-contrast T1-weighted, T2-weighted, and Flair. Collectively, the complementary information from the aforementioned modalities enables a more robust segmentation of a tumor brain.

Recently, deep learning methods have been successfully applied in a variety of application domains~\cite{krizhevsky2012imagenet, girshick2014rich, samadi2019change}. One of the applications is brain tumor segmentation which has gained wide attention in the medical imaging field. In this paper, we investigate some of the best recent methods and techniques adopted for automatic brain tumor segmentation task.

Havaei et al.~\cite{havaei2017brain} introduced specific multi-path CNNs to segment brain tumor regions over 2D slices of MRI images. Additionally, two training phases were used to tackle the imbalanced classes of the input data. A boundary-aware FCN was developed by Shen et al.~\cite{shen2017boundary} to increase the segmentation performance. Subsequently, Kamnitsas et al.~\cite{kamnitsas2017efficient} proposed a novel 3D network, named Deep Medic, which extracts multi-scale feature maps, and then integrates them locally and globally by using a two-path architecture.

According to the effectiveness of encoder-decoder architectures for the purpose of semantic segmentation, networks like UNet~\cite{ronneberger2015u} also appropriately performed the brain tumor segmentation. All the winners who participated in BRAin Tumor Segmentation 2017 (BRATS 2017) challenge also benefited from encoder-decoder networks. Applying anisotropic convolutions, Wang et al.~\cite{wang2017automatic} trained three networks for every tumor sub-region in the cascade form, where the previous network output was considered as the subsequent network input.

\begin{figure*}[!tp]
\begin{center}
\includegraphics[width=\textwidth]{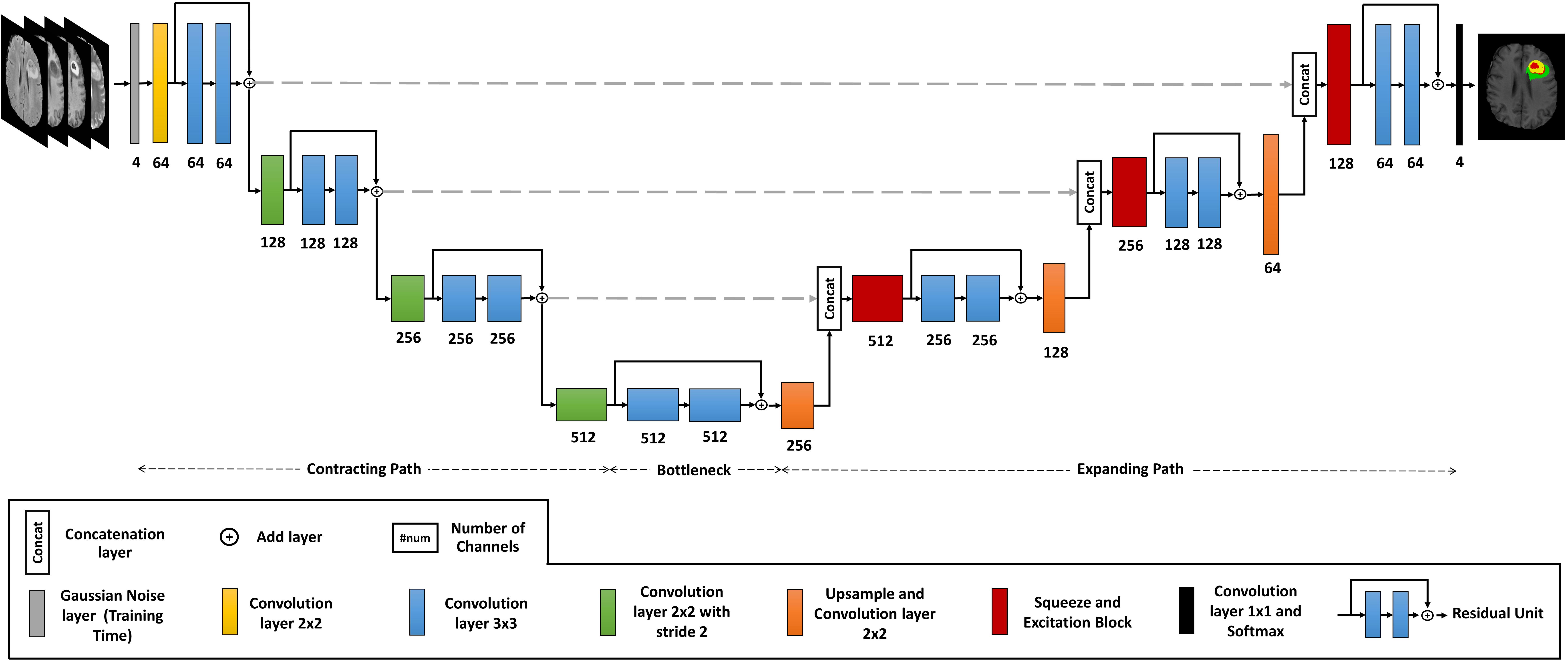}
\end{center}
   \caption{The architecture of our proposed network which is based on UNet architecture.}
\label{fig:Main}
\end{figure*}

In BRATS 2018, Andriy Myronenko~\cite{myronenko20183d}, Isensee et al.~\cite{isensee2018no}, and McKinley et al.~\cite{mckinley2018ensembles} ranked first, second, and third, respectively. Andriy Myronenko~\cite{myronenko20183d} indicated that the presence of a variational auto-encoder branch for the construction of an input image could provide the capability of regularizing the shared decoder. Isensee et al.~\cite{isensee2018no} used a minor modification version of the UNet architecture. Moreover, the authors utilized extra training data provided by their own institution to improve the overall performance. McKinley et al.~\cite{mckinley2018ensembles} developed a shallow network by applying a structure in which the dilated convolution was employed. Furthermore, to calculate the uncertainty of the label, the generalization of binary Cross-entropy was used. Zhou et al.~\cite{zhou2018learning} suggested the application of an ensemble of various networks incorporating multi-scale contextual information, adding an attention block, and segmentation of three tumor sub-regions in the cascade form. Wang et al.~\cite{wang2018automatic} investigated Test Time Augmentation technique (TTA) in which various augmentation methods are applied at the test time. By adopting TTA on several networks, they demonstrated that it can improve the overall performance of brain tumor segmentation.

Several studies~\cite{cciccek20163d, isensee2017brain, baid2018deep} have shown that the 3D versions of UNet architecture are able to achieve better results compared to fully 2D architectures. Although 3D UNet has good performance, it has more parameters and computational complexity than 2D version and that is why we used a version of 2D UNet architecture to enhance the performance of the network in terms of memory. Consequently, we need to extract 2D slices from 3D volumes of MRI images, which causes not to benefit from 3D contextual information of input images. To overcome this problem, we have employed the Multi-View technique to enhance the network performance by benefiting from 3D contextual information of input images. Moreover, since most of the recent methods (especially UNet-based networks) integrate low-level and high-level features in a naive way, i.e. considering equal importance for each feature map, it may result in confusion for the model. To address this problem, we propose an extended version of UNet architecture in which we adopted channel attention mechanism technique after concatenation of low-level and high-level features by weighting each channel adaptively. 

In summary, our paper carries three contributions as follows: 
\begin{itemize}
\item Since we used a version of 2D UNet architecture due to containing low number of  parameters, we employ Multi-View Fusion technique to benefit from 3D contextual information of input images to improve the performance. 
\item Integrating multi-level features in a naive way may result in confusion for the model. Thus, we propose an extended version of UNet architecture in which we adopted the channel attention mechanism after concatenation of multi-level features by weighting each channel adaptively to prevent confusion for the model. 
\item Although our method is 2D, it performs favorably against 2017 and 2018 state-of-the-art methods. 

\end{itemize}

 \section{Method}
 \subsection{Network}
 
Our proposed network is based on UNet structure~\cite{ronneberger2015u} with some modifications, which is illustrated in Fig.~\ref{fig:Main}. The network consists of two main paths: i) the contracting path that encodes the whole input image and ii) the expanding path that recovers the original resolution. 
The contracting path has three layers of downsampling in which, instead of using Max-pooling layers employed in the original UNet, we utilize convolutional layers with Stride = 2 similar to~\cite{milletari2016v}. 
 
We also adopt residual units, as in~\cite{he2016identity,he2016deep}, instead of plain units in the original UNet to speedup training and convergence. In each residual unit, there are two convolutional layers each of which followed by a Batch Normalization layer~\cite{ioffe2015batch} and PReLU activation~\cite{he2015delving} instead of ReLU activation used in the original UNet. After the contracting path, a residual unit is used in the bottleneck to connect both paths. Similarly, three residual units are used in the expanding path. This path also has three upsampling layers each of which doubles the size of feature maps. Moreover, a $2 \times 2$ convolutional layer is adopted after each upsampling layer.

\begin{figure}[!t]
\begin{center}
\includegraphics[width=\linewidth]{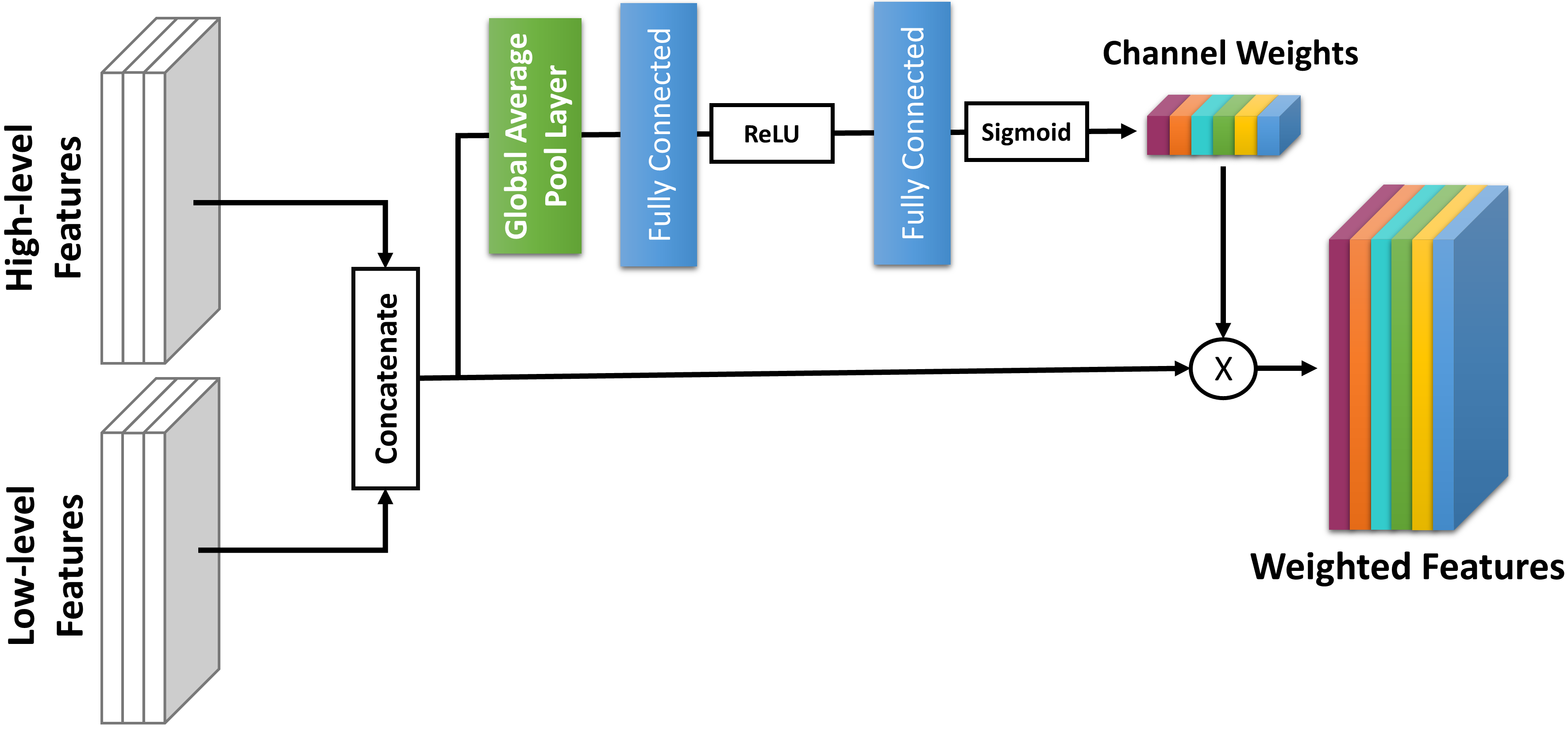}
\end{center}
   \caption{Illustration of the attention mechanism in which an SE Block~\cite{hu2018squeeze} is adopted on concatenated multi-level features}
\label{fig:SE}
\end{figure}

To combine the feature maps in the expanding path with the corresponding feature maps of the contracting path, the original UNet used the direct concatenation. However, the direct concatenation of these high-level and low-level feature maps without weighting their importance is not the best way to efficiently integrate them. As a matter of fact, the multi-level features may not be useful for all types of input images and this would lead to redundancy of information. Moreover, inaccurate and ambiguous information of some levels might cause confusion for the network, and thus leads to wrong segmentation of tumors. 
To address these problems, different from the original UNet, we utilize a channel attention mechanism by adopting a Squeeze and Excitation Block~\cite{hu2018squeeze} after each concatenation layer to adaptively weight the channels, as illustrated in Fig.~\ref{fig:SE}. This design generates channel weights to re-weight the concatenated feature maps. Finally, at the end of the network, a $1 \times 1$ convolutional layer followed by a Softmax function is adopted to map the features of the previous layer to the expected number of classes.

\subsection{Loss Function}
The overall performance of a segmentation model depends not only on the architecture of the network but also on the choice of the loss function~\cite{sudre2017generalised}, particularly in the cases that suffer from highly class imbalance problems. Therefore, choosing an appropriate loss function becomes more challenging. Due to the distributions of the tumor and non-tumor regions, the brain tumor segmentation task has an innate class imbalance problem. 
Thus, the widely-used loss functions in the segmentation tasks are not appropriate for training our network. If these functions are adopted, the network tries to learn the larger classes and this results in poor segmentation performance. 
To tackle this problem, we use Generalized Dice loss (GDL)~\cite{sudre2017generalised}, which adaptively weights the classes to balance them, along with the well-known Cross-entropy loss (CE), which speeds up the convergence.

\begin{equation}
L_{overall}=L_{GDL}(G,P)+ \lambda \times L_{CE}(G,P)
\end{equation}
 
\noindent where the $\lambda$ is empirically set to $1.25$. 

The GDL function~\cite{sudre2017generalised} is a multi-class version of the Dice loss function. GDL also assigns an adaptive weight to each class to deal with the imbalances of brain tumor classes. GDL is computed as: 

\begin{equation}
L_{GDL}(G,P)= 1-2  \frac{\sum_{j=1}^C \big( W_{j}\times \sum_{i=1}^N (g_{ij}\times p_{ij})\big)+  \epsilon}{\sum_{j=1}^C \big( W_{j}\times \sum_{i=1}^N (g_{ij}+ p_{ij})\big)+  \epsilon }
\end{equation}

\noindent where $\epsilon$ is regularization constant and $W_{j}$ is the adaptive weight for $j-th$ class and formulated as:

\begin{equation}
W_j = \frac{1}{(\sum_{i=1}^N g_{ij})}
\end{equation}

The multi-class Cross-entropy loss function is also computed as follows: 

\begin{equation}
L_{CE}(G,P)=-\frac{1}{N}\sum_{i=1}^N \sum_{j=1}^C \big(g_{ij} \times Log(p_{ij})\big)
\end{equation}

\subsection{Multi-view Fusion}
\begin{figure}[!b]
\begin{center}
\includegraphics[width=\linewidth]{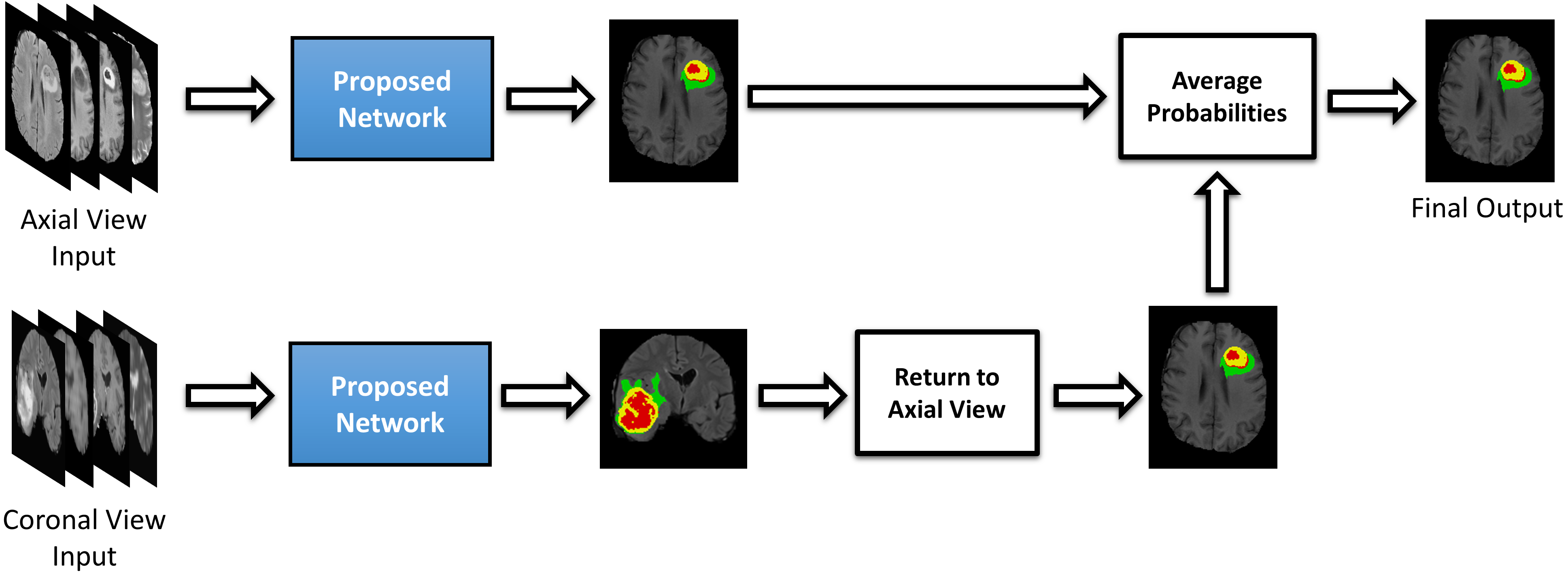}
\end{center}
   \caption{Multi-view Fusion process.}
\label{fig:MultiView}
\end{figure}

Since our proposed network is a 2D architecture, we need to extract 2D slices from 3D volumes of MRI images. To benefit from 3D contextual information of input images, we extract 2D slices from both Axial and Coronal views, and then train a network for each view separately. In the test time, we build the 3D output volume for each model by concatenating the 2D predicted maps. Finally, we fuse the two views by pixel-wise averaging. The whole procedure is illustrated in the Fig.~\ref{fig:MultiView}. Benefiting from the fusion of these two views, our method is capable of considering the 3D nature of input images. We show the effectiveness of the Multi-view Fusion in the experimental results section. 

\section{Experiments}
\subsection{Dataset description}
In this paper, we utilized the BRATS 2017 and 2018~\cite{menze2014multimodal, bakas2017advancing} datasets for experiments. The training sets of these datasets contain 3D MRI volumes of 285 different patients, in which there are 210 volumes as HGG and 75 volumes as LGG with dimensions of $240 \times 240 \times 155$. The BRATS 2017 and the BRATS 2018 validation sets contain 3D MRI volume of 46 and 66 patients of unknown grades, respectively. There are four modalities for each individual brain, namely, T1, T1c (post-contrast T1), T2, and Flair which were skull-stripped, resampled and coregistered. These datasets includes four labels, namely, enhancing tumor, edema, necrosis, and background. For the purpose of evaluation, annotations are merged into three binary sub-regions including Whole Tumor region, Tumor Core region, and Enhancing Tumor region, which we denote them as WT, TC, and ET, respectively. The experts of this domain have created ground truth by manual segmentation. The segmentation labels for the validation sets are not publicly available, and the participants must upload the results provided by their networks to the BRATS online evaluation platform in order to obtain quantitative evaluations such as Dice Score and Hausdorff Distance~\cite{taha2015metrics}.

\subsection{Evaluation metrics}
To evaluate the performance of our method, We use Dice Score and Hausdorff Distance for ET, WT, and TC sub-regions.

Dice Score calculates the similarity between X and Y sets as follows:

\begin{equation}
Dice =\frac{2\mid X\cap Y \mid}{\mid X\mid + \mid Y \mid}
\end{equation}

\noindent where $\mid X \mid$ and $\mid Y \mid$ denote the cardinalities of $X$ and $Y$ sets, respectively.

The Hausdorff Distance (HD) is defined as the longest distance between a point in a one set and the most adjacent point of the other set and defined as:
\begin{multline}
HD(X,Y)=max \big\{ sup_{x\in X}\ \ inf_{y\in Y} \ \ d(x,y),\\ sup_{y\in Y}\ \ inf_{x\in X} \ \ d(x,y) \big\}
\end{multline}

\noindent where d(x,y) is Euclidean Distance between ${x\in X}$ and ${y\in Y}$. To reduce problems with noisy predictions, $95th$ percentile is used instead of the max operation, which we refer to as Hausdorff95.

\subsection{Implementation Details}
We conduct all of our experiments in Google Colabatory service. Our proposed network is developed in Keras~\cite{chollet2015keras}, using TensorFlow~\cite{tensorflow2015-whitepaper} backend. For each view, we train our network by performing a 5-fold cross-validation on the 285 cases of BRATS 2018 training set (228 cases for training and the other 57 cases for validation for each fold). Finally, evaluation results for each view are obtained by adopting an ensemble learning method in which we average the Softmax output of the five networks. BRATS 2018 and 2017 validation sets are used to evaluation our method and all reported results were computed by the online evaluation platform. Our final results are available in the leaderboard section of these challenges under the title "IUST\_ ICCKE2019".

For pre-processing the data, firstly, N4ITK algorithm, a bias field correction algorithm proposed in~\cite{tustison2010n4itk}, is adopted on each MRI modalities to correct the inhomogeneity of these images. Secondly, $1\%$ of the top and bottom intensities is removed like~\cite{havaei2017brain}, and then each modality is normalized to zero mean and unit variance. To reduce overfitting, two kinds of data augmentations are employed at random: vertical and horizontal flipping. A Gaussian noise layer with standard deviation 0.01 is also utilized at the input of architecture in other to tackle the noisy nature of MRI images. To train the networks, we adopt SGD~\cite{bottou2010large} with a momentum 0.9 and learning rate of 8e-3.

\begin{table}[!b]
\centering
\caption{The ablation analysis of our proposed network. Models are trained on the first fold and the Axial view. The best results are shown in {\color[HTML]{FE0000} \textbf{red}}.}
\label{tab:abl}
\resizebox{\linewidth}{!}{%
{\renewcommand{\arraystretch}{1.4}
\begin{tabular}{l|ccc}
\hline
                                                                                                                & \multicolumn{3}{c}{Dice}                                                                                             \\ \cline{2-4} 
\multirow{-2}{*}{Network}                                                                                       & ET                                    & WT                                    & TC                                    \\ \hline
\textbf{UNet}                                                                                                  & 0.751                                 & 0.879                                 & 0.791                                 \\
\textbf{UNet + Minor Modifications}                                                                            & 0.768                                 & 0.883                                 & 0.807                                 \\ \hline
\textbf{\begin{tabular}[l]{@{}l@{}}UNet + Minor Modifications + Attention Mechanism \\(the proposed network)\end{tabular}} & {\color[HTML]{FE0000} \textbf{0.776}} & {\color[HTML]{FE0000} \textbf{0.888}} & {\color[HTML]{FE0000} \textbf{0.821}} \\ \hline
\end{tabular}%
}
}
\end{table}

\begin{table}[!b]
\centering
\caption{The 5-fold cross-validation ensemble results on Axial and Coronal views along with the results of the Multi-view Fusion technique. The best results are shown in {\color[HTML]{FE0000} \textbf{red}}.}
\label{tab:EnsView}
\resizebox{\linewidth}{!}{%
{\renewcommand{\arraystretch}{1.4}
\begin{tabular}{l|ccc|ccc}
\hline
                                                                                & \multicolumn{3}{c|}{Dice}                                                                                             & \multicolumn{3}{c}{Hausdorff95}                                                                                   \\ \cline{2-7} 
\multirow{-2}{*}{Network}                                                       & ET                                    & WT                                    & TC                                    & ET                                   & WT                                   & TC                                   \\ \hline
\textbf{\begin{tabular}[c]{@{}l@{}}Axial View 5-fold Ensemble\end{tabular}}   & 0.793                                 & {\color[HTML]{FE0000} \textbf{0.895}} & {\color[HTML]{FE0000} \textbf{0.826}} & 4.69                                 & 4.30                                 & 9.54                                 \\
\textbf{\begin{tabular}[c]{@{}l@{}}Coronal View 5-fold Ensemble\end{tabular}} & 0.810                                 & 0.886                                 & 0.811                                 & 3.25                                 & 5.54                                 & 8.38                                 \\ \hline
\textbf{\begin{tabular}[c]{@{}l@{}}Multi-view Fusion\end{tabular}}            & {\color[HTML]{FE0000} \textbf{0.813}} & {\color[HTML]{FE0000} \textbf{0.895}} & {\color[HTML]{000000} 0.823}          & {\color[HTML]{FE0000} \textbf{2.93}} & {\color[HTML]{FE0000} \textbf{4.05}} & {\color[HTML]{FE0000} \textbf{6.34}} \\ \hline
\end{tabular}%
}
}
\end{table}

\subsection{Ablation Study}
In this section, we show the effectiveness of the modifications employed in the proposed network. The modifications are divided into two parts, the Minor Modifications (including Residual Units,  strided convolution, PReLU and BN) and the attention mechanism which weights the multi-level features by adopting SE Blocks. For the sake of simplicity, all the models in this section are trained on the first fold and the Axial view of the BRATS 2018 training set, and then evaluated on the BRATS 2018 validation set. As seen from Table~\ref{tab:abl}, the attention mechanism improves the overall performance significantly which shows the beneficial effect of weighting multi-level features.

\begin{figure*}[htp] 

\centering
\subfigure[a HGG example (subject name: BRATS18\_TCIA08\_436\_1).]{\includegraphics[scale=0.47]{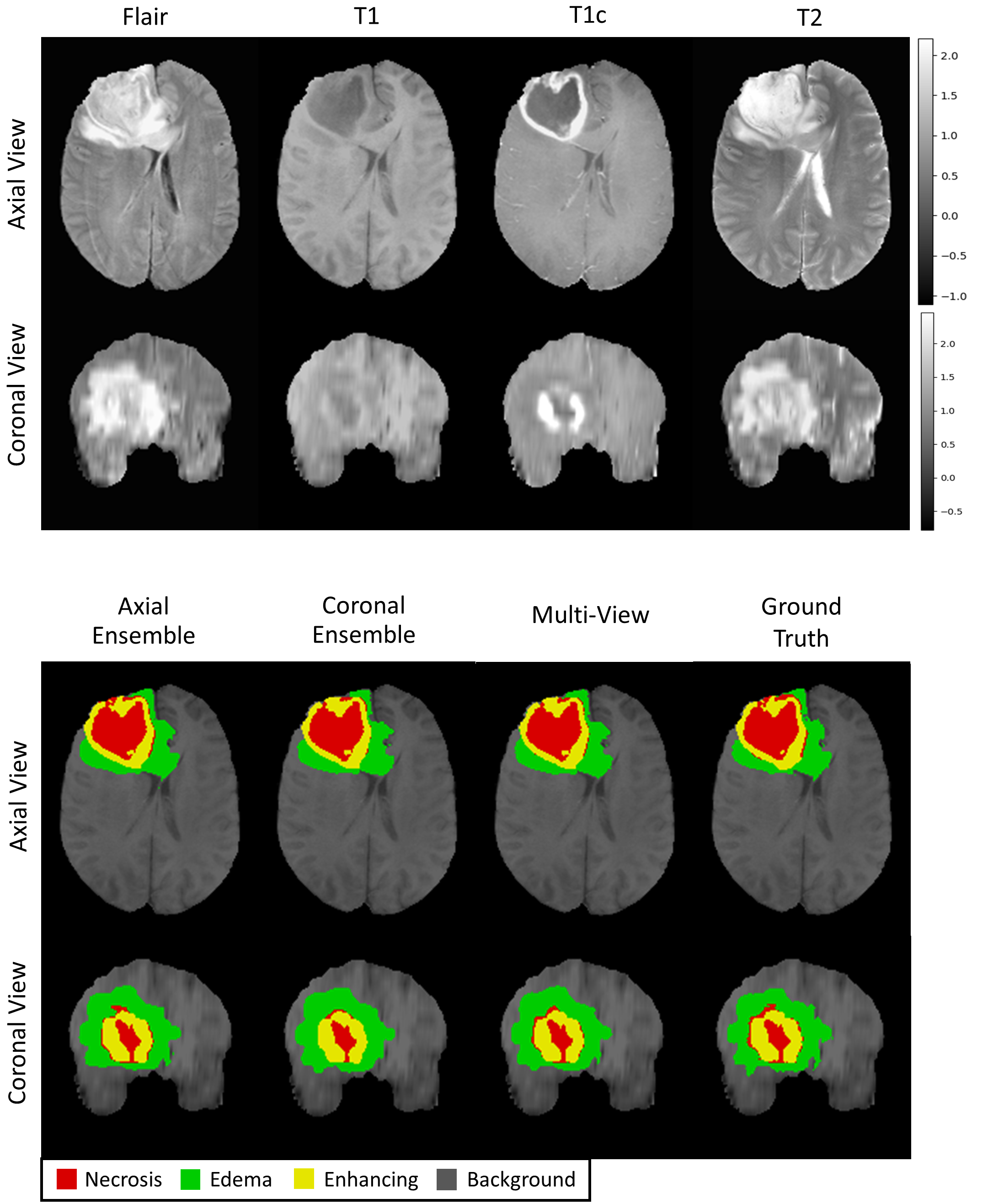}}\quad
\subfigure[a LGG example (subject name: BRATS18\_TCIA12\_298\_1).]{\includegraphics[scale=0.47]{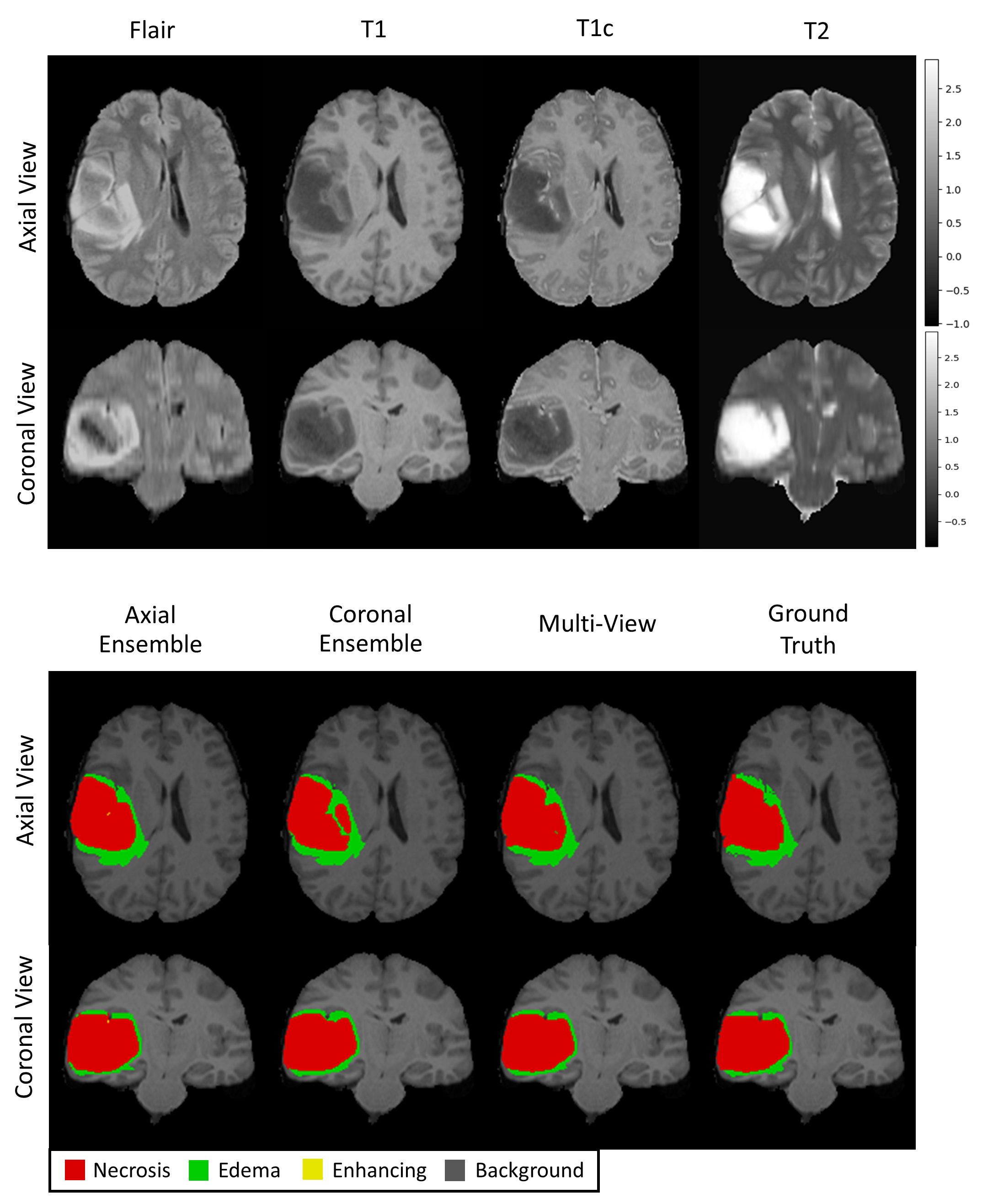}}

\caption{Visual comparison examples of the 5-fold cross-validation ensemble results on Axial and Coronal views along with the results of the Multi-view Fusion for a HGG tumor and a LGG tumor.} 
\label{fig:HGGLGG}
\end{figure*}

\begin{figure*}[htp]
\begin{center}
\includegraphics[width=0.9\textwidth]{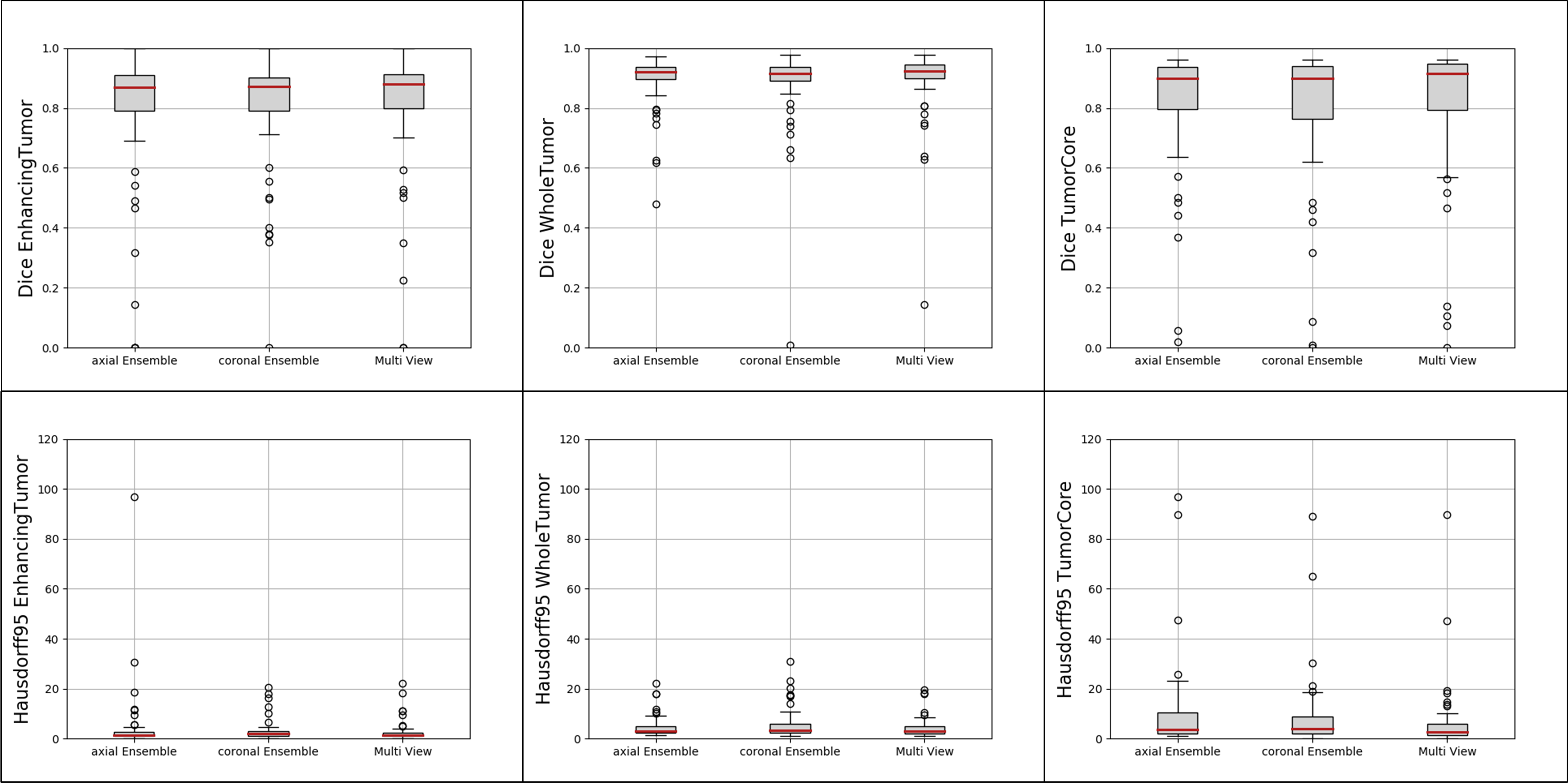}
\end{center}
   \caption{Box plots of the 5-fold cross-validation ensemble results on Axial and Coronal views along with the results of the Multi-view Fusion.}
\label{fig:Box}
\end{figure*}

\subsection{The Proposed Method Results}
In this section, evaluation results of our proposed method are reported. Table~\ref{tab:EnsView} demonstrates the ensemble results of 5-fold cross-validation on Axial and Coronal views along with the results of the Multi-view Fusion technique. As seen from Table~\ref{tab:EnsView}, Multi-view Fusion improves the overall performance, especially in terms of Hausdorff distance metric, which shows the beneficial effect of considering the 3D nature of the data. The Fig.~\ref{fig:HGGLGG} shows the visual comparison of the networks in Table~\ref{tab:EnsView} for a HGG tumor and a LGG tumor. In this figure, the input slices and the segmentation masks are shown in Axial view and Coronal view. We also provide Dice score and Hausdorff95 Box plots in Fig.~\ref{fig:Box} for the three regions.

\subsection{Comparison with the Existing Methods}
In this section, the proposed method is evaluated on the validation dataset of the BRATS 2017 and BRATS 2018 challenges, and compared with the best methods of these challenges in terms of Dice score and Hausdorff distance metric. As previously mentioned, all results are obtained from the online evaluation platform. Table~\ref{tab:Brats2018} and Table~\ref{tab:Brats2017} show the evaluation results of BRATS 2018 and 2017 validation set, respectively. Although our network is a 2D architecture, it performs favorably against state-of-the-art methods, especially in terms of ET Dice score and Hausdorff95 in all three sub-regions.

\begin{table}[]
\centering
\caption{Comparison of our method and the methods of BRATS 2018 on validation set. {\color[HTML]{FE0000} \textbf{red}} and {\color[HTML]{3166FF} \textbf{blue}} demonstrates the best two results, respectively.}
\label{tab:Brats2018}
\resizebox{\linewidth}{!}{%
{\renewcommand{\arraystretch}{1.4}
\begin{tabular}{l|c|ccc|ccc}
\hline
                         &                                    & \multicolumn{3}{c|}{Dice}                                                                                             & \multicolumn{3}{c}{Hausdorff95}                                                                                   \\ \cline{3-8} 
\multirow{-2}{*}{Method} & \multirow{-2}{*}{Type}                                                            & ET                                    & WT                                    & TC                                    & ET                                   & WT                                   & TC                                   \\ \hline
\textbf{Myronenko}~\cite{myronenko20183d}       & 3D                                 & {\color[HTML]{FE0000} \textbf{0.823}} & {\color[HTML]{FE0000} \textbf{0.910}} & {\color[HTML]{FE0000} \textbf{0.867}} & 3.93                                 & {\color[HTML]{333333} 4.52}          & {\color[HTML]{333333} 6.85}          \\
\textbf{Isensee et al.}~\cite{isensee2018no}  & 3D                                 & 0.804                                 & {\color[HTML]{3166FF} \textbf{0.908}} & {\color[HTML]{3166FF} \textbf{0.854}} & {\color[HTML]{000000} 3.12}          & 4.97                                 & {\color[HTML]{000000} 7.04}          \\
\textbf{McKinley et al.}~\cite{mckinley2018ensembles}  & 3D                                 & 0.796                                 & 0.903                                 & 0.847                                 & 3.55                                 & {\color[HTML]{3166FF} \textbf{4.17}} & {\color[HTML]{FE0000} \textbf{4.93}} \\
\textbf{Zhou et al.}~\cite{zhou2018learning}     & 3D                                 & 0.792                                 & 0.907                                 & 0.835                                 & {\color[HTML]{FE0000} \textbf{2.80}} & {\color[HTML]{333333} 4.48}          & 7.07                                 \\
\textbf{Gholami et al.}~\cite{gholami2018novel}    & 3D                                 & 0.791                                 & 0.908                                 & 0.819 & ---                                  & ---                                  & ---                                  \\
\textbf{Albiol et al.}~\cite{albiol2018extending}      & 3D                                 & 0.773                                 & 0.881                                 & 0.777 & ---                                  & ---                                  & ---                                  \\
\textbf{Chen et al.}~\cite{chen2018s3d}        & 3D                                 & 0.733                                 & 0.888                                 & 0.808                                 & 4.64                                 & 5.51                                 & 8.14                                 \\ \hline
\textbf{Ours}        & {\color[HTML]{FE0000} \textbf{2D}} & {\color[HTML]{3166FF} \textbf{0.813}} & 0.895                                 & 0.823                                 & {\color[HTML]{3166FF} \textbf{2.93}} & {\color[HTML]{FE0000} \textbf{4.05}} & {\color[HTML]{3166FF} \textbf{6.34}} \\ \hline
\end{tabular}%
}
}
\end{table}

\begin{table}[]
\centering
\caption{Comparison of our method and the methods of BRATS 2017 on validation set. {\color[HTML]{FE0000} \textbf{red}} and {\color[HTML]{3166FF} \textbf{blue}} demonstrates the best two results, respectively.}
\label{tab:Brats2017}
\resizebox{\linewidth}{!}{%
{\renewcommand{\arraystretch}{1.4}
\begin{tabular}{l|c|ccc|ccc}
\hline 
                                                                                &                                    & \multicolumn{3}{c|}{Dice}                                                                                             & \multicolumn{3}{c}{Hausdorff95}                                                                                   \\ \cline{3-8} 
\multirow{-2}{*}{Method}                                                        & \multirow{-2}{*}{Type}                               & ET                                    & WT                                    & TC                                    & ET                                   & WT                                   & TC                                   \\ \hline
{\color[HTML]{000000} \textbf{Wang et al.}~\cite{wang2017automatic}}           & 3D                                 & {\color[HTML]{3166FF} \textbf{0.786}} & {\color[HTML]{FE0000} \textbf{0.905}} & {\color[HTML]{FE0000} \textbf{0.838}} & {\color[HTML]{FE0000} \textbf{3.28}} & {\color[HTML]{FE0000} \textbf{3.89}} & {\color[HTML]{FE0000} \textbf{6.48}} \\
{\color[HTML]{000000} \textbf{Kamnitsas et al.}~\cite{kamnitsas2017ensembles}} & 3D                                 & {\color[HTML]{333333} 0.757}          & {\color[HTML]{3166FF} \textbf{0.902}} & {\color[HTML]{3166FF} \textbf{ 0.820}}          & {\color[HTML]{333333} 4.22}          & {\color[HTML]{333333} 4.56}          & {\color[HTML]{3166FF} \textbf{6.11}} \\
{\color[HTML]{000000} \textbf{Isensee et al.}~\cite{isensee2017brain}}         & 3D                                 & {\color[HTML]{333333} 0.732}          & {\color[HTML]{333333} 0.896}          & {\color[HTML]{333333} 0.797}          & {\color[HTML]{333333} 4.55}          & {\color[HTML]{333333} 6.97}          & {\color[HTML]{333333} 9.48}          \\

{\color[HTML]{000000} \textbf{Feng et al.}~\cite{feng2017patch}}              & 3D                                 & {\color[HTML]{333333} 0.751}          & {\color[HTML]{333333} 0.896}          & {\color[HTML]{333333} 0.799}          & {\color[HTML]{333333} 4.76}          & {\color[HTML]{333333} 12.53}         & {\color[HTML]{333333} 8.69}          \\
{\color[HTML]{000000} \textbf{Jesson et al.}~\cite{jesson2017brain}}                                          & 3D                                 & {\color[HTML]{333333} 0.713}          & {\color[HTML]{333333} 0.899}          & {\color[HTML]{333333} 0.751} & {\color[HTML]{333333} 6.98}  & {\color[HTML]{333333} 4.16}  & {\color[HTML]{333333} 8.65}  \\

{\color[HTML]{000000} \textbf{Andermatt et al.}~\cite{andermatt2017automated}} & 3D                                 & {\color[HTML]{333333} 0.711}          & {\color[HTML]{333333} 0.893}          & {\color[HTML]{333333} 0.734}          & {\color[HTML]{333333} 4.19}          & {\color[HTML]{333333} 4.61}          & {\color[HTML]{333333} 8.19}          \\ 
{\color[HTML]{000000} \textbf{Jungo et al.}~\cite{jungo2017towards}}          & {\color[HTML]{FE0000} \textbf{2D}} & {\color[HTML]{333333} 0.674}          & {\color[HTML]{333333} 0.884}          & {\color[HTML]{333333} 0.726}          & {\color[HTML]{333333} 6.63}          & {\color[HTML]{333333} 7.93}          & {\color[HTML]{333333} 10.91}         \\ \hline

{\color[HTML]{000000} \textbf{Ours}}                                        & {\color[HTML]{FE0000} \textbf{2D}} & {\color[HTML]{FE0000} \textbf{0.791}} & {\color[HTML]{333333} 0.885}          & {\color[HTML]{333333} 0.783}          & {\color[HTML]{3166FF} \textbf{3.36}} & {\color[HTML]{3166FF} \textbf{4.11}} & {\color[HTML]{333333} 7.36}          \\ \hline
\end{tabular}%
}
}
\end{table}

\section{Conclusion}
In this paper, we propose an improved version of 2D UNet architecture for the purpose of segmenting brain tumors. Despite designing a 2D architecture that contains the low number of parameters, the model can benefit from 3D contextual information of input images by using the Multi-View technique. Moreover, since considering equal importance for each feature map after concatenation of low-level and high-level features may result in confusion for the model, we utilize the attention mechanism to extract discriminative features and prevent confusion for the model. By adopting these techniques, we achieve average Dice scores of 0.813, 0.895, and 0.823 for ET, WT, and TC, respectively, by using the BRATS 2018 validation data set.



%

%
%
%
%

\bibliographystyle{IEEEtran}

\bibliography{article_bib}

\end{document}